\documentclass[aps]{revtex4}
\usepackage{amssymb}
\usepackage{bm}

\begin{document}

\title{Theory of spin-Hall
transport of heavy holes in semiconductor quantum wells}

\author{P.~Kleinert\footnote[3]{e-mail: kl@pdi-berlin.de},
V.V.~Bryksin\dag{}, }

\affiliation{Paul-Drude-Institut f\"ur Festk\"orperelektronik,
Hausvogteiplatz 5-7, 10117 Berlin, Germany} \affiliation{\dag\
Physical Technical Institute, Politekhnicheskaya 26, 194021 St.
Petersburg, Russia}

\begin{abstract}
Based on a proper definition of the spin current, we investigate
the spin-Hall effect of heavy holes in narrow quantum wells in the
presence of Rashba spin-orbit coupling by using a spin-density
matrix approach. In contrast to previous results obtained on the
basis of the conventional definition of the spin current, we
arrive at the conclusion that an electric-field-induced
steady-state spin-Hall current does not exist in both, pure and
disordered infinite samples. Only an ac field can induce a
spin-Hall effect in such systems.
\end{abstract}


%\submitto{\JPCM}

\maketitle

\section{Introduction}
\label{intro} The emerging field of spintronics stimulates
extensive studies of the spin-Hall effect (SHE) in semiconductors
with spin-orbit coupling. Recently, an intrinsic SHE, which is
entirely due to spin-orbit coupling, has been predicted to appear
in two-dimensional (2D) electron systems \cite{Sinova1} and in
$p$-doped bulk semiconductors \cite{Murakami}. It has been argued
that this effect occurs even in the absence of any scattering
events and that the spin-Hall conductivity is given by a universal
constant. Subsequently, a detailed treatment of elastic scattering
revealed that scattering effects can lead to a complete
cancellation of the total intrinsic spin-Hall current
\cite{Inoue1,Mishchenko,Dimitrova,Khaetskii,Liu73}. Especially the
simple Rashba model of a 2D electron gas with its unusual
properties gave rise to many controversies. Other models that have
been studied seem to be more robust against impurity scattering
\cite{Nomura1}. This discussion is nowadays replaced by a more
fundamental debate about the proper definition of the spin current
\cite{Rashba2,Zhang94,PZhang,Sugimoto,Brykspin,Sun}. Due to
non-conservation of spin, which results from spin precession, it
has been argued that its definition is largely a matter of
convenience \cite{Shytov}. In analogy to the charge current, most
researchers identified the spin current simply with the
expectation value of the product of spin and velocity observables.
Based on this definition, the spin current is expressed from a
technical point of view by particular elements of the spin-density
matrix namely
$f_{\lambda^{\prime}}^{\lambda}({\bm{k}},{\bm{\kappa}}\mid
t)\mid_{\bm{\kappa}=\bm{0}}$ (with the spin quantum numbers
$\lambda$, $\lambda^{\prime}$, the time variable $t$, and the
quasi-momenta ${\bm{k}}=({\bm{k}}_1+{\bm{k}}_2)/2$,
${\bm{\kappa}}={\bm{k}}_1-{\bm{k}}_2$), which depend only on the
quasi-momentum ${\bm{k}}$. These quantities are conventionally
calculated from multiband Boltzmann equations. However, this
procedure is not sufficiently general and even fails for
particular charge transport problems. There is a subtlety related
to this definition of transport coefficients. Already for the
charge transport, this procedure is only justified, when the
interaction Hamiltonian of the system commutes with the position
operator. Otherwise one has to go back to the more general
definition, which expresses the charge current through the time
derivative of the dipole operator \cite{alt1,alt2}. In this case,
the current of the multiband system is not obtained from the
special elements of the density matrix
$f_{\lambda^{\prime}}^{\lambda}({\bm{k}}\mid t)$, but is
calculated from quantities
$\nabla_{\bm{\kappa}}f_{\lambda^{\prime}}^{\lambda}({\bm{k}},{\bm{\kappa}}\mid
t)\mid_{\bm{\kappa}=\bm{0}}$, which requires the consideration of
the ${\bm{\kappa}}$ dependence in kinetic equations. For charge
transport, the theory of small polarons provides a famous example.
In accordance with these experiences in the charge transport
theory, a proper definition of the spin current has recently been
put forward \cite{PZhang}. According to this proper definition,
the spin current is expressed by the time derivative of the spin
displacement. This concept of spin current has a number of
remarkable advantages \cite{PZhang}. It is in accordance with the
near-equilibrium transport theory, satisfies the Onsager
relations, and provides vanishing spin currents for Anderson
insulators. As a result, by determining conjugate forces,
thermodynamic and electric measurements of the spin current become
possible. To further illustrate the advantage of the proper
definition, let us consider a particular feature of the spin
current. Due to its symmetry with respect to time inversion, a
finite spin current can exist even in equilibrium \cite{Rashba2}.
For simplicity we treat the Rashba model of a 2D electron gas.
According to the conventional definition, there is a
field-independent stationary spin-Hall current that is completely
independent of the spin accumulation. In contrast, the spin-Hall
current according to the proper definition turns out to be due to
the initial time variation of the spin polarization and disappears
in the steady state, when the spin polarization becomes constant
\cite{Brykspin}.

In the present work, we study the SHE for heavy holes in III-V
semiconductor quantum wells on the basis of the proper definition
of the spin current by systematically deriving and solving the
kinetic equations for the spin-density matrix. Recently,
experimental observations of the SHE have been reported for such a
system \cite{Wunderlich}. The analysis \cite{Nomura1} of the
experimental results seems to provide evidence for a close
correspondence between the detected edge spin accumulation and the
bulk spin currents as described by the conventional spin Hall
theory. However, this conclusion cannot be considered to be final
as the spin polarization near the edges depends on the boundary
conditions and may not be directly induced by spin-Hall currents
far from the boundaries. The treatment of inhomogeneous and/or
finite samples requires the calculation of
$f_{\lambda^{\prime}}^{\lambda}({\bm{k}},{\bm{\kappa}}\mid t)$,
the ${\bm{\kappa}}$ dependence of which introduces a strong
coupling between the spin and charge degrees of freedom even in
the absence of an electric field. The theory of spin transport in
finite samples has its particular own challenges.

\section{Theory}
We focus on a model for the lowest heavy hole subbands, which is
described by a one-particle Hamiltonian for heavy holes in narrow
quantum wells being subject to spin-orbit interaction of the
Rashba type that results from structural inversion asymmetry. The
Hamiltonian of the 2D hole gas in the second quantized form
\begin{eqnarray}
H&=&\sum_{\bm{k},\lambda }a_{\bm{k}\lambda }^{\dag}\left[ \varepsilon_{%
\bm{k}}-\varepsilon _{F}\right] a_{%
\bm{k}\lambda }-\sum_{\bm{k},\lambda ,\lambda ^{\prime }}\left(
\hbar\vec{\bm{\omega}}_{
\bm{k}} \cdot \vec{\bm{\sigma }}_{\lambda \lambda ^{\prime }}\right) a_{\bm{k}%
\lambda }^{\dag}a_{\bm{k}\lambda ^{\prime }}\nonumber\\
&-&ie\vec{\bm{E}} \sum_{\bm{k},\lambda}\left. \nabla_{\bm \kappa}
a^{\dag}_{\bm{k}-\frac{\bm{\kappa}}{2}\lambda}a_{\bm{k}+
\frac{\bm{\kappa}}{2}\lambda}\right|_{\bm{\kappa}=\bm{0}}+
u\sum\limits_{{\bm{k}},{\bm{k}}^{\prime}}
\sum\limits_{\lambda}a_{{\bm{k}}\lambda}^{\dag}a_{{\bm{k}}^{\prime}\lambda},
\label{Hamil}
\end{eqnarray}
is composed of the creation ($a^{\dag}_{\bm{k}\lambda}$) and
annihilation ($a_{\bm{k}\lambda}$) operators with quasi-momentum
${\bm{k}}=(k_x,k_y,0)$ and spin $\lambda$. The dispersion relation
of free in-plane motion is given by
$\varepsilon_{\bm{k}}=\hbar^2k^2/2m$. Throughout, the Fermi energy
$\varepsilon_F$ is always assumed to be the largest relevant
energy scale. The strength of the "{}white-noise"{} disorder
scattering is given by the scattering rate $\hbar/\tau=2\pi
u\rho_F$ with $\rho_F$ and $u$ denoting the 2D density of states
in the absence of spin-orbit coupling and the scattering strength,
respectively. For simplicity, the electric field ${\bm{E}}$ is
oriented along the $x$ axis. The cubic Rashba spin-orbit coupling
\cite{Gerchikov,Schlie71,Liu72} is obtained from the Pauli
matrices and the energy
\begin{equation}
\hbar\vec{\bm{\omega}}_{\bm{k}}=\frac{\alpha}{2}\left[i(k_{+}^3-k_{-}^3),(k_{+}^3+k_{-}^3),0\right],
\label{omega}
\end{equation}
with $k_{\pm}=k_{x}\pm i k_{y}$ and $\hbar\omega_{k}=\alpha k^3$.
$\alpha$ denotes the spin-orbit coupling constant.

All information that is necessary to determine the kinetic
observables are contained in the spin-density matrix, the elements
of which are grouped in the following form
\begin{equation}
f(\bm{k},{\bm{\kappa}}\mid
t)=\sum\limits_{\lambda}f_{\lambda}^{\lambda}(\bm{k},{\bm{\kappa}}\mid
t),\quad \vec{\bm{f}}(\bm{k},{\bm{\kappa}}\mid
t)=\sum\limits_{\lambda,\lambda^{\prime}}
f_{\lambda^{\prime}}^{\lambda}(\bm{k},{\bm{\kappa}}\mid
t)\vec{\bm{\sigma}}_{\lambda\lambda^{\prime}}.%
\label{densitydef}
\end{equation}
Although we treat a homogeneous 2D infinite hole gas, it is
indispensable to retain both the $\bm{k}$ and $\bm{\kappa}$
dependence of the spin-density matrix in order to calculate the
proper spin current. In the thermodynamic equilibrium
($\bm{E}=\bm{0}$) and to lowest order in $\alpha$, the spin degree
of freedom is characterized by the vector
\begin{equation}
\vec{\bm{f}}_{eq}(\bm{k}\mid t)=-\hbar\vec{\bm{\omega}}_{\bm{k}}
\frac{\partial n(\varepsilon_{\bm{k}})}{\partial
\varepsilon_{\bm{k}}},
\end{equation}
with $n(\varepsilon_{\bm{k}})$ being two times the Fermi
distribution function. With respect to the time dependence, we
prefer the consideration of Laplace-transformed kinetic equations,
which are derived by exploiting the Born approximation and by
keeping only the lowest-order contributions of the spin-orbit
interaction in the collision integral. Due to the $\bm{\kappa}$
dependence, the resulting kinetic equations
\begin{equation}
sf-\frac{i\hbar}{m}(\bm{\kappa}\cdot\bm{k})f+i\vec{\bm{\omega}}_{\bm{\kappa}}
({\bm{k}})\cdot\vec{\bm{f}} +\frac{e\vec{\bm{
E}}}{\hbar}\nabla_{\bm{k}}f
=\frac{1}{\tau}(\overline{f}-f)+n(\varepsilon_{\bm{k}}),
\label{kin1}
\end{equation}
\begin{eqnarray}
&&s\vec{\bm{f}}+2(\vec{\bm{\omega}}_{\bm{k}}\times\vec{\bm{f}})
-\frac{i\hbar}{m}(\bm{\kappa}\cdot\bm{k})\vec{\bm{f}}
+i\vec{\bm{\omega}}_{\bm{\kappa}} ({\bm{k}})f+\frac{e\vec{\bm{
E}}}{\hbar}\nabla_{\bm{k}}\vec{\bm{f}}\nonumber\\
&&=\frac{1}{\tau}(\overline{\vec{\bm{f}}}-\vec{\bm{f}})+\frac{1}{\tau}
\frac{\partial}{\partial\varepsilon_{\bm{k}}}
\overline{f\hbar\vec{\bm{\omega}}_{\bm{k}}}-\frac{\hbar\vec{\bm{\omega}}_{\bm{k}}}{\tau}
\frac{\partial}{\partial\varepsilon_{\bm{k}}} \overline{f}
-\hbar\vec{\bm{\omega}}_{\bm{k}}\frac{\partial
n(\varepsilon_{\bm{k}})}{\partial \varepsilon_{\bm{k}}},
\label{kin2}
\end{eqnarray}
are coupled to each other. An integration over the angle of the
vector $\bm{k}$ is indicated by a bar over the respective
quantity. In addition, we make use of the abbreviation
\begin{equation}
\hbar\vec{\bm{\omega}}_{\bm{\kappa}}({\bm{k}})=3\alpha
\left[(k_y^2-k_x^2)\kappa_y-2k_xk_y\kappa_x,(k_x^2-k_y^2)\kappa_x-2k_xk_y\kappa_y,0
\right].
\end{equation}
Next, the spin current is treated. According to Refs.
\cite{PZhang,Brykspin}, the proper spin current is calculated from
the derivative of the spin displacement
\begin{equation}
\vec{\bm{j}}_{i}(t)=\frac{3\hbar}{2}\frac{\partial}{\partial t}
\sum\limits_{\bm{m}}\sum\limits_{\lambda_1,\lambda_2}
{\bm{r}}_{\bm{m}i}\langle
a_{\bm{m}\lambda_1}^{\dag}a_{\bm{m}\lambda_2}\rangle_{t}
\vec{\bm{\sigma}}_{\lambda_1\lambda_2},
\end{equation}
where ${\bm{r}}_{\bm{m}}$ is the position operator and the factor
$3\hbar/2$ refers to the angular momentum of the heavy holes.
Performing a Fourier transformation and an integration by parts,
the Laplace transformed spin-Hall current is expressed by the
equivalent equation
\begin{equation}
j_y^z(s)=-is\frac{3\hbar}{2}\left. \sum\limits_{\bm{k}}
\frac{\partial}{\partial
\kappa_y}{\vec{\bm{f}}^{z}}({\bm{k}},{\bm{\kappa}}\mid
s)\right|_{\bm{\kappa}=\bm{0}}. \label{sdef}
\end{equation}
Within the linear response approach, we need only the first-order
corrections of the spin-density matrix with respect to the
electric field $\bm{E}$ and the quasi-momentum $\bm{\kappa}$.
Based on perturbation theory, analytical results for this quantity
are derived. Details of the calculation are given in the Appendix.

To start our analysis, let us first treat the field-induced spin
accumulation. From the analytical solution in Eq.~(A.4) together
with (A.5) and (A.6), we conclude that there is no field-induced
spin accumulation in the cubic Rashba model
($\overline{\vec{\bm{f}}}_{\bm{0}\bm{E}}={\bm{0}}$). This result
should be compared with the finite current-induced spin
polarization in the linear Rashba model \cite{Edelstein}.
Furthermore, in contrast to the linear Rashba model there is no
spin current, which is independent of the electric field and which
exists even in thermodynamic equilibrium. Both peculiarities of
the cubic Rashba model are closely related to each other and are
compatible with recent studies based on diffusion equations
\cite{Bleineu}.

To calculate the spin-Hall current, we use the analytical results
from the Appendix and obtain for its nonzero component an
expression
\begin{equation}
j_y^z(s)=-27\frac{\alpha^2}{\hbar^2}eE\sigma\sum\limits_{\bm{k}}
n(\varepsilon_{\bm{k}})k^4\frac{\sigma^2-4\omega_k^2}{(\sigma^2+4\omega_k^2)^3},
\label{st1}
\end{equation}
which indicates that all occupied states below the Fermi energy
contribute. The very same feature, which is also observed in the
linear Rashba model, seems to be a generic property of the
intrinsic SHE \cite{Brykspin,Liu72}. In Eq.~(\ref{st1}), the
abbreviation $\sigma=s+1/\tau$ is used. Integrating by parts, we
obtain another equivalent result, in which at zero temperature
($T=0$) only states on the Fermi surface play a role
\begin{equation}
j_y^z(s)=\frac{9}{2}\frac{eE}{m}\sigma\sum\limits_{\bm{k}}
n^{\prime}(\varepsilon_{\bm{k}})\frac{\omega_k^2}{(\sigma^2+4\omega_k^2)^2}.
\end{equation}
The time dependence of the spin-Hall current is obtained from this
equation by applying an inverse Laplace transformation
\begin{equation}
j_y^z(t)=\frac{9}{8}\frac{eE}{m}\exp{\left(-\frac{t}{\tau}\right)}
\sum\limits_{\bm{k}}
n^{\prime}(\varepsilon_{\bm{k}})(\omega_kt)\sin (\omega_kt).
\end{equation}
After the electric field is switched on at time $t=0$, $j_y^z(t)$
exhibits damped oscillations. In the steady state, there is no
spin-Hall current. At zero temperature, our general result takes
the simple form
\begin{equation}
j_y^z(s)=-\frac{9}{2\pi}\frac{eE\alpha^2}{\hbar^2} \frac{\sigma
k_{F}^6}{(\sigma^2+4\alpha^2k_{F}^6)^2}, \label{st2}
\end{equation}
where $k_F$ denotes the Fermi momentum. From this equation, we
obtain for the spin-Hall conductivity of a perfect crystal
($\tau\rightarrow\infty$) the result
\begin{equation}
\sigma_{sH}(\omega\rightarrow 0)=\frac{9e}{8\pi\hbar^2}
\left(\frac{\omega}{2\omega_{k_F}}\right)^2,\label{st3}
\end{equation}
which vanishes in the limit $\omega\rightarrow 0$. This conclusion
completely contradicts previous findings derived on the basis of
the conventional definition of the spin current \cite{Schlie71}.
We regain this published result for the spin-Hall current from
Eqs.~(A.4) to (A.6). From the general expression for $T=0$
\begin{equation}
j_y^z(s)=-\frac{9}{2\pi}\frac{eE\alpha^2}{\hbar^2 s}
\frac{k_{F}^6}{\sigma^2+4\alpha^2k_{F}^6},
\end{equation}
the universal spin-Hall conductivity
\begin{equation}
\sigma_{sH}(\omega\rightarrow 0)=-\frac{9e}{8\pi\hbar^2},
\end{equation}
is obtained. These results, which have previously been derived by
an alternative approach \cite{Schlie71}, strongly deviate from
Eqs.~(\ref{st2}) and (\ref{st3}). According to the widespread
reasoning, the SHE is present and robust against disorder in the
cubic Rashba model. In contrast, based on the proper definition of
the spin current, we come to the conclusion that the SHE is absent
in an infinite heavy hole gas.

\section{Summary}
Recently, it has been recognized that the SHE has an intrinsic
contribution due to spin-orbit interaction in a perfect crystal.
This assertion has occupied a great deal of attention because of
its potential for electronic devices with low power consumption. A
theoretical controversy about the disorder effect that can
completely eliminate the intrinsic SHE seems to be settled now:
the intrinsic SHE is absent only in the simple model of a 2D
electron gas with Rashba spin-orbit coupling that turns out to
exhibit anomalous properties. In all other generic systems that
have been specifically studied, the SHE is present and robust
against disorder \cite{Nomura1}. However, this conclusion is
derived on the basis of an approach for the SHE that revealed
rather unconventional properties
\cite{Rashba2,Zhang94,PZhang,Sugimoto,Brykspin,Sun} so that
serious doubts arose on its physical relevance. A recent proper
definition of the spin current \cite{PZhang,Brykspin} resolves a
number of difficulties of former approaches and is in line with
Onsager relations that allow the application of the
near-equilibrium transport theory. On the basis of this physically
motivated concept of spin transport, we arrive at the conclusion
that neither in the linear \cite{Brykspin} nor in the cubic Rashba
model an intrinsic SHE exists. Only an ac electric field is able
to induce a spin-Hall current. This conclusion does not contradict
recent experiments \cite{Wunderlich}, which revealed edge spin
accumulations. The observed spin polarization near the boundaries
certainly depends to a large extend on the boundary conditions and
may not be simply due to spin Hall currents in the bulk. In an
inhomogeneous system, there is always a strong coupling between
the charge and spin degrees of freedom that can give rise to a
number of generic effects not present in an infinite homogeneous
2D electron or hole gas. It is an interesting future task to
extend the approach applicable for the bulk to the spin-transport
phenomena in finite systems.
\section*{Acknowledgements}

Partial financial support by the Deutsche Forschungsgemeinschaft
and the Russian Foundation of Basic Research under the grant
number 05-02-04004 is gratefully acknowledged.

\appendix
\section{Solution of kinetic equations}
The proper definition of the spin current in Eq.~(\ref{sdef}) sets
up our calculational scheme as an perturbational approach with
respect to $\bm{\kappa}$. In addition, we restrict ourselves to
the linear response regime, where only the lowest-order
contribution in the electric field is taken into account. The
formal solution of the kinetic Eq.~(\ref{kin2}) has the form
\begin{equation}
\vec{\bm{f}}=\frac{\sigma\bm{X}-2\,
\vec{\bm{\omega}}_{\bm{k}}\times\bm{X} +4\,
\vec{\bm{\omega}}_{\bm{k}}(\vec{\bm{\omega}}_{\bm{k}}\cdot
\bm{X})/\sigma}{\sigma^2+4\omega_{\bm{k}}^2}, \label{los1}
\end{equation}
where $\sigma=s+1/\tau$. $\bm{X}$ comprises all angle integrated
quantities as well as the $\bm{\kappa}$ and field contributions.
This result can be used to execute the perturbational approach
step by step. First, we treat the equations for
$\bm{\kappa}=\bm{0}$ and $\bm{E}=\bm{0}$ and obtain immediately
\begin{equation}
f_{00}=\frac{n(\varepsilon_{\bm{k}})}{s}=\overline{f_{00}},\quad
\vec{\bm{f}}_{00}=-\hbar\vec{\bm{\omega}}_{\bm{k}}\frac{n^{\prime}}{s},\quad
\overline{\vec{\bm{f}}_{00}}=\bm{0}, \label{f00}
\end{equation}
where the indices $00$ refer to the order in $\bm{\kappa}$ and
$\bm{E}$. $n^{\prime}$ is a short-hand notation for $\partial
n(\varepsilon_{\bm{k}})/\partial\varepsilon_{\bm{k}}$. Next, the
lowest-order correction due to the electric field is calculated
($\bm{E}\ne\bm{0}$, $\bm{\kappa}=\bm{0}$). The kinetic
Eq.~(\ref{kin1}) for the charge degree of freedom is easily solved
\begin{equation}
f_{0\bm{E}}=-\frac{eE}{\sigma s}\frac{\hbar k_x}{m}n^{\prime},
\quad \overline{f_{0\bm{E}}}=0 .\label{f0E}
\end{equation}
Based on Eq.~(A.1), we obtain for the spin contribution
\begin{equation}
\vec{\bm{f}}_{0\bm{E}}=\frac{\sigma\bm{R}_{0\bm{E}}-2\,
\vec{\bm{\omega}}_{\bm{k}}\times\bm{R}_{0\bm{E}} +4\,
\vec{\bm{\omega}}_{\bm{k}}(\vec{\bm{\omega}}_{\bm{k}}\cdot
\bm{R}_{0\bm{E}})/\sigma}{\sigma^2+4\omega_{\bm{k}}^2},
\label{los1}
\end{equation}
with
\begin{equation}
\bm{R}_{0\bm{E}}^{x}=\alpha\frac{eE}{\hbar s}\left[
n^{\prime\prime}\frac{\hbar^2k_xk_y}{m}(k_y^2-3k_x^2)
-6n^{\prime}k_xk_y\right],
\end{equation}
\begin{equation}
\bm{R}_{0\bm{E}}^{y}=\alpha\frac{eE}{\hbar s}\left[
n^{\prime\prime}\frac{\hbar^2k_x^2}{m}(k_x^2-3k_y^2)
+3n^{\prime}(k_x^2-k_y^2)\right],\quad \bm{R}_{0\bm{E}}^{z}=0.
\end{equation}
From this solution, we conclude that there is no field-induced
spin accumulation $\overline{\vec{\bm{f}}}_{\bm{0}\bm{E}}=\bm{0}$.
This peculiarity is specific for the cubic Rashba model. Within
the framework of the conventional theory, all components of the
spin-density matrix that determine the spin transport have already
been obtained. However, for the treatment of the proper spin
current it is not sufficient to calculate $\vec{\bm{f}}(\bm{k}\mid
s)$, it is rather necessary to treat
$\nabla_{\bm{\kappa}}\vec{\bm{f}}(\bm{k},\bm{\kappa}\mid s)$ at
$\bm{\kappa}=\bm{0}$. Therefore, we have to extend the
perturbational approach with respect to $\bm{\kappa}$. For
$\bm{\kappa}\ne\bm{0}$ and $\bm{E}=\bm{0}$, we obtain
\begin{equation}
f_{\bm{\kappa}0}=({\bm{\kappa}}\cdot{\bm{k}}) \left[\frac{i\hbar
n}{m\sigma s}+3i\frac{\alpha^2n^{\prime}}{\hbar\sigma
s}k^4\right],
\end{equation}
\begin{equation}
\vec{\bm{f}}_{\bm{\kappa}0}=-\frac{i\hbar^2}{m}\frac{n^{\prime}}{\sigma
s}({\bm{\kappa}}\cdot{\bm{k}})\vec{\bm{\omega}}_{\bm{k}}
-\frac{in}{s}\frac{\sigma\vec{\bm{\omega}}_{\bm{\kappa}}
-2\vec{\bm{\omega}}_{\bm{k}}\times\vec{\bm{\omega}}_{\bm{\kappa}}
+4\vec{\bm{\omega}}_{\bm{k}}(\vec{\bm{\omega}}_{\bm{k}}\cdot
\vec{\bm{\omega}}_{\bm{\kappa}})/\sigma}{\sigma^2+4\omega_k^2},
\end{equation}
where the second term on the right-hand side of Eq.~(A.7) can be
neglected. Finally, we compute the quantity that determines the
spin-Hall current in Eq.~(\ref{sdef}), namely the contribution,
which is proportional to $\bm{\kappa}$ and $\bm{E}$
\begin{equation}
\overline{\vec{\bm{f}}_{\bm{\kappa}\bm{E}}^{z}}=\frac{\sigma
\overline{\vec{\bm{R}}_{\bm{\kappa}\bm{E}}^{z}}
-2\overline{(\vec{\bm{\omega}}_{\bm{k}}\times
\vec{\bm{R}}_{\bm{\kappa}\bm{E}})^{z}}}{\sigma^2+4\omega_k^2},
\end{equation}
with
\begin{equation}
\vec{\bm{R}}_{\bm{\kappa}\bm{E}}=\frac{i\hbar}{m}({\bm{\kappa}}\cdot{\bm{k}})
\vec{\bm{f}}_{0\bm{E}}-i\vec{\bm{\omega}}_{\bm{\kappa}}({\bm{k}})
f_{0\bm{E}}-\frac{eE}{\hbar}\frac{\partial}{\partial
k_x}\vec{\bm{f}}_{\bm{\kappa}0}+\frac{1}{\tau}
\frac{\partial}{\partial\varepsilon_{\bm{k}}}
\overline{\hbar\vec{\bm{\omega}}_{\bm{k}}f_{\bm{\kappa}\bm{E}}}.
\end{equation}
After a lengthy but straightforward calculation, we obtain the
result for the spin-Hall current in Eq.~(\ref{st1}) from
Eqs.~(A.9), (A.10), and (\ref{sdef}).

\section*{References}
% BibTeX users please use
%%\bibliographystyle{prsty}
%%\bibliography{abbrev,spin}

%
\end{document}